\documentclass[aps,prl,twocolumn,showpacs,psfig]{crtex}
\usepackage{psfig}
\usepackage{amssymb}
\include{bibliohead}

\def\ber{\begin{eqnarray}}
\def\eer{\end{eqnarray}}
\def\bea{\begin{equation}}
\def\eea{\end{equation}}
\def\lya{Ly$\alpha$ }

\begin{document}
\preprint{}
\title{Can Cosmic Rays Heat the Intergalactic Medium? }
\author{Saumyadip Samui, Kandaswamy Subramanian, Raghunathan Srianand}
\affiliation{IUCAA, Post Bag 4, Ganeshkhind, Pune - 411 007, India}
\email{samui@iucaa.ernet.in, kandu@iucaa.ernet.in, anand@iucaa.ernet.in}
\date{}
\begin{abstract}

Supernova explosions in the early star forming galaxies will accelerate
cosmic rays (CRs). CRs are typically 
confined in the collapsed objects for a short period before escaping 
into the intergalactic medium (IGM). 
Galactic outflows can facilitate this escape by advecting CRs into
the IGM. An outflow that results in a termination shock can also 
generate more CRs. We show that the CR protons from 
the above processes 
can significantly affect the thermal history of the IGM.
Within plausible range of parameters, cosmic ray heating can 
compensate for adiabatic cooling and explain 
the measured IGM temperature at redshifts $z \sim 2-4$, 
even with early reionization. 
\end{abstract}
\maketitle
It is widely believed that the temperature of the
low density intergalactic medium retains some 
memory of the reionization process \cite{igmT}.
Analysis of \lya absorption spectra in the framework of hydrodynamical
simulations and semi-analytic models \cite{schaye} suggest that, 
the mean temperature of the IGM is in the range $10^4\le T({\rm K})\le
3\times10^4$ for 2 $\le z\le 4$. This has been used 
to argue that hydrogen reionization
occurred below $z\simeq 9$ \cite{tom}. Had the reionization
occurred at higher $z$, adiabatic cooling would have brought the
temperature much below the measured value at $z = 4$. The
detection of a strong Gunn-Peterson trough  and the sizes of 
ionized regions around the highest $z$ QSOs \cite{fan}, 
are consistent with a large neutral hydrogen fraction at $z\simeq6$.
The CMB data from the WMAP satellite \cite{kogut}, 
however, points to an earlier 
epoch of reionization in the range $11<z<30$. Thus 
either new heating sources (other than photo-heating) 
or multiple episodes of reionization with 
heating at low $z$ provided by, say, the He~II reionization induced 
by QSOs \cite{hui_haiman}, 
are needed to explain the IGM temperature at $z \sim 2- 4$. 
Here, we consider the effect of cosmic ray protons on the thermal history
of the IGM. 

Cosmic rays are expected to be generated and accelerated
from the shocks created by the
exploding supernovae (SNe) in the star forming regions, 
termination shocks created by outflows from galaxies and 
accretion shocks during structure formation.
The properties of CRs in our Galaxy is well documented 
\cite{schlickeiser,ber,dogiel,hillas}.
The energy density in the proton component is 
about $1 $ eV cm$^{-3}$, and CRs are thought to be confined 
to the Galactic disk for about $10^7$ yr before 
escaping, presumably into the
IGM. It is widely accepted that $\ge 15\%$ of the average
SNe energy must go into accelerating the CRs so that the flux
density of CRs can be maintained at the observed value \cite{hillas}.
Low energy CR protons
play an important role in the ionization and thermal state of
the gas in the interstellar medium (ISM) of normal galaxies. 
The possibility that they could also play
an important role in determining the thermal history of the
IGM has been considered sporadically in the 
literature \cite{ginzburg_ozernoy,biman}.
The existence of good observational constraints on (i) the 
global star formation rate (SFR) density as a function of $z$
(ii) epoch of reionization  and (iii) temperature of the IGM
makes it appropriate to re-examine this issue using a model that
self consistently computes the generation and spectral evolution 
of CRs in the IGM. We do this here in the framework of 
a standard flat $\Lambda$CDM model ($\Omega_m=0.27$,
$\Omega_b =  0.04$ and $H_0 =70~{\rm km~s}^{-1}~{\rm Mpc}^{-1}$ ).

\par

Consider CR protons produced from the SNe explosions.
To compute the rate of energy injected into CR protons
from early star forming galaxies, one needs to know the 
SFR and the number of SNe 
explosions, $f_{SN}$, per solar mass of forming stars.
We use the observed global comoving SFR 
density \cite{bouwens}, $R(z)$ 
measured in units of $M_{\odot}$ yr$^{-1}$ Mpc$^{-3}$ and approximated by
\ber
R (z) = 7.5\times10^{-4} \eta~ z^{2.5} e^{-0.8 ( z - 3.5 ) } .
\label {SFR}
\eer
Here, $\eta$, in the range 5 to 9,
is the unknown scaling factor used to correct
the effect of dust reddening \cite{adelberger}.
We shall adopt $\eta=7$ in our models.
Further, using the ``Starburst99'' code \cite{leitherer}
we get $f_{SN} = 8\times10^{-3}$ 
for Salpeter IMF (with M$_{min}$ = 0.1 M$_\odot$,
M$_{max}$ = 100 M$_\odot$)
and $f_{SN} = 1.7\times 10^{-2} $ for a top-hat mass function
with M$_{min}$ = 40 M$_\odot$, M$_{max}$ = 100 M$_\odot$
(denoted as ``top-heavy''). 
We assume that the energy released from 
a single SNe is $ 10^{51} {\rm erg}$ and a fraction $\epsilon=0.15 $ of this 
is utilized for accelerating the CR protons.
The average rate of energy injected in the CR protons is then,
\begin{equation}
\dot E_{CR}(z)
 = 10^{-30}\epsilon \ R \ f_{SN} (1+z)^3~{\rm erg}~{\rm s}^{-1}~{\rm cm}^{-3}.
\end{equation}
\par\noindent
The cosmic ray proton spectrum from the SNe is in general taken to be
a power law in momentum space, with $ n( p )\; dp \propto p^{-q} dp$, 
with $q=2.2$ as a typical value \cite{schlickeiser}.
We can then compute the average rate of CRs generated in  
star forming regions as  
\bea
 \frac{dn(p,z)}{dt}\; dp = \dot N_0 p^{-q} dp 
\eea
where the normalization $\dot N_0$ for a given $z$ is fixed by integrating 
$E(p)\dot N_0 p^{-q} dp$ using a lower energy cut-off \cite{schlickeiser} 
of 10 keV, and equating the result 
to $\dot E_{CR}(z)$. Here $E(p)$ is  
the kinetic energy of a CR proton corresponding to a momentum $p$.
This injected spectrum will be modified as the CRs propagate
first within the high redshift star forming galaxy and then into
the IGM. The evolution of the velocity $v=\beta c$ of the
CR particle due to collisions with free electrons, hydrogen ionization,
and the adiabatic expansion of the universe is given by \cite{schlickeiser,biman},
\ber
-\frac { d \beta } { dt } & = & 3.27 \times 10 ^{ -16 } n_e ( z )
\frac { ( 1 - \beta ^2 ) ^{3/2} } { \beta } \frac { \beta ^2 } { x_m ^3 + \beta ^3 } \nonumber \\
& & + ~ 1.94 \times 10^{-16} n_{HI}(z) \frac{(1-\beta ^2)^{3/2}}{\beta } \nonumber \\
& & \times \frac{ 2 \beta ^2 }{[ ( 0.01 )^3 + 2 \beta ^3 ]} [1 + 0.0185
\ln \beta ~ \Theta ( \beta - 0.01) ] \nonumber \\
& & ~~~~~~- \frac { \beta ( 1 - \beta ^2 ) } { 1 + z } \frac { dz } { dt }.
\label{dbetadt}
\eer
Here, $n_e$ and $n_{HI}$ are the densities
of electrons and neutral hydrogen respectively in units of 
cm$^{-3}$. $\Theta$ is Heaviside function,
$ x_m = 0.0286~( T / 2 \times 10 ^6 K)^{1/2} $ 
and $T$ is the temperature of the gas. 
The energy losses are 
dominated by low energy CRs for $\beta>x_m$.
Typically a CR particle of energy E
will lose all its energy to the electrons in a time, $t_L \simeq
1.5\times10^8 (E/3 {\rm MeV})^{1.5}(10^{-4}/n_e)$ yr, after traveling a
distance $d_L\sim 4 (E/3 {\rm MeV})^2(10^{-4}/n_e)$ Mpc. 
\begin{figure}[]
\centerline{\vbox{
\psfig{figure=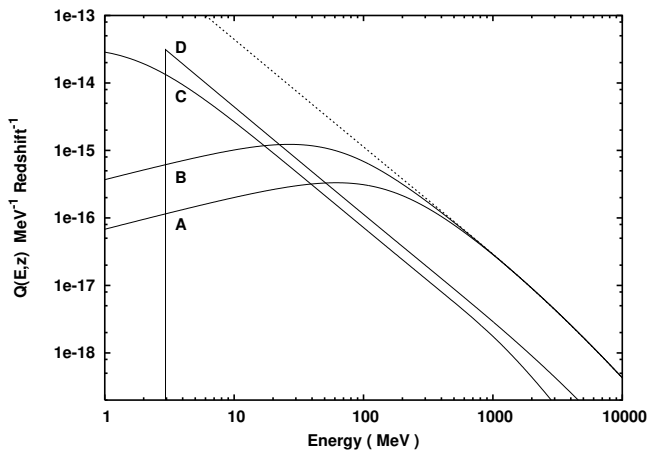,width=8.6 cm,angle=-90.}}}
\caption[]{Cosmic rays spectrum originating from SNe (dotted one) 
and the modified spectra (solid lines) escaping from galaxy.
The curves A and B are for $x = 10$ and $3$ g cm$^{-2}$ respectively.
The curve C is for the advection model, where the spectrum A has
been modified by decreasing the  momentum of each CR particle 
by a factor 10. Curve D is the CR spectrum originating
from the termination shock with a sharp cut-off at 3 MeV.
For illustrative purpose the injected $Q(E,z)$ is 
normalized using $R$ at $z = 15$.}
\label{spectra}
\end{figure}

The escape of the CRs from the star forming galaxies 
will depend on the structure and strength of its magnetic 
field, the gas density ($\rho$) and outflows etc. 
We consider three possible scenarios for the emerging CR spectrum:
(I) CRs are confined for a period $t_c$ to the galaxy
and then escape into the IGM,
(II) after a confinement time $t_c$,
CRs are advected by galactic outflows into the IGM, 
and (III) CRs are generated in termination shocks
created by the outflows.

In case (I), we compute the emergent spectrum from the galaxy, taking
into account the energy loss to electrons, as a function of grammage, 
$x\sim\rho c t_{c}$~g cm$^{-2}$, traversed by the CRs within 
the galaxy. For our Galaxy $x$ is $\sim10$ g cm$^{-2}$
\cite{dogiel}. For the high $z$ galaxies, $x$ could be smaller,
because the global star formation rate density is expected
to be dominated by smaller mass objects that may also be 
`magnetically' younger and will confine CRs for a smaller period.
The injected spectrum as well as modified
spectra (defined as the rate $Q(E,z)dE = (dn/dt)(dt/dz) dp$ per unit
redshift) for  $x =10,~3$ g cm$^{-2}$,
are shown in the Fig.  \ref{spectra}
and labeled as A and B.
Larger and larger part of the low energy CR-spectrum is cut off 
due to energy losses as one increases $x$. As an example,
when $x = 10$ g cm$^{-2}$ protons of initial energy $\le 160$ MeV
will lose all their energy. 
Energy loss from protons with higher energy does
lead to an universal low energy tail with $Q \propto E^{1/2}$.
Since $dE/dt \propto E^{-1/2}$, the heating is still dominated
by protons near the peak of $Q$.

High redshift Lyman break galaxies (LBGs) show signatures of strong
outflows \cite{shapley}. Such outflows, that are also 
required to explain the 
observed IGM metallicity, will help in advecting the
CRs into the IGM. The adiabatic loss in such outflows 
can also lower the momentum ($p$) of the CR particles, 
with $p\propto\rho^{1/3}$, thereby shifting the spectrum to  
lower momentum and decreasing the total energy in CRs.
Suppose a wind from a galaxy of 3 kpc size expands up to 30 kpc
before CRs escape into the IGM. The average gas density will decrease by
a factor 10$^3$ there by reducing the momentum of CRs by a factor 10.
The emergent spectrum for this case (our case (II))
(and $x=10$ g cm$^{-2}$) 
is also shown in Fig. \ref{spectra} (see curve C).

In order for the CRs to influence the properties of the IGM,
they should be released into the IGM within the Hubble time. 
This time-scale is confinement time for case (I) and advection
time for case (II). 
In our Galaxy $t_c\sim10^{7}$ yr in the disk. 
In the Galactic halo of say a radius $r_H$, 
the typical $t_c$ is the diffusion time, 
$r_H^2/6D \sim 10^8~{\rm yr}$, 
for $r_H\sim10$ kpc and a diffusion coefficient 
$D = 5\times 10^{28}$ cm$^2$ s$^{-1}$ \cite{ber}. 
Thus for case (I) $t_c$  is expected to be smaller than the Hubble time, 
even for the Galaxy. Similarly the advection time scale in
an outflow of velocity
typically seen in LBGs (i.e $\sim$100 km s$^{-1}$) 
to a radius 30 kpc as
in our case (II) will be 3 $\times 10^{8}$ yr, which
is shorter than the Hubble time for $z\le15$.

In case (III) we consider the CRs produced from the termination 
shocks of the outflows. We assume that about $10\%$ of the 
SNe energy is channeled into driving such an outflow, and 
a fraction $\epsilon$ of this goes into accelerating the CRs in
the resulting termination shock. The spectrum of CRs emerging
from the shock is again expected to be a power law in momentum
space. As the wind termination shock is expected to form 
far out in the halo of the galaxy and propagate 
for most periods in the IGM itself, even low energy
cosmic rays can be directly injected into the IGM.
However the lowest energy CRs will lose their energy
in the vicinity of the galaxy itself. For example, 
a 3 MeV CR proton will travel a proper distance $d_L \sim 10$ Mpc
in the IGM at $z \sim 5$ before losing all its energy. While
within a halo that has 200 times larger density, it will
travel $\sim 50$ kpc. Thus we assume that the particles with 
energy $ E \geq 3$ MeV
will be able to escape from termination shocks and contribute
to the heating of the IGM. The resulting spectrum with
a sharp cut-off at $3$ MeV 
is shown in Fig.  \ref{spectra} as case D.

The free propagation of CRs in the IGM will depend on the strength
and coherence of the intergalactic magnetic field.
For example in the case of ISM the streaming of CRs is limited by 
resonant scattering on self-generated Alfv\'en waves.
It turns out that a coherent IGM magnetic field larger than 
$\sim 10^{-14}$ G is required for a 
streaming instability growth 
time-scale $< 10^{8}$ yr (Eq.~32 in \cite{zweibel}),
so that there can be adequate growth of Alfv\'en waves, by
say $z \sim 5$. 
Astrophysical battery mechanisms produce only 
tiny coherent IGM seed fields 
$\sim 10^{-23} - 10^{-19}$ G \cite{subetal}.
Early Universe generation mechanisms are not very compelling
because the results are often exponentially sensitive to 
parameters involved \cite{widrow}. Further, for $z\ge2$, the 
self-confinement due to plasma waves generated by the 
streaming CRs \cite{ginzburg_ozernoy} is inefficient  
because of the strong damping of waves by electron ion 
collisions. Here, we will assume that the CRs escaping 
the high $z$ galaxies are not self confined, but can 
stream out freely in the IGM. 

The spectrum of cosmic ray background at a given $z$ is the sum
of CR spectrum injected at $z$ and that evolving from higher redshifts.
The number density of CRs, $N_{CR}(\beta , z, z_0 )d\beta$, 
in an interval ($\beta$, $\beta+d \beta$), at any given epoch, $z$,
is 
\bea
N_{CR}( \beta , z, z_0 ) =  \int\limits_{z_0}^z dz_i
\frac { dn(z_i, p _i) } { dz_i }\frac { dp_i } { d \beta_i } 
\frac { d\beta _i } { d \beta }
\left( \frac { 1 + z } { 1 + z_i } \right) ^3. 
\label {NCR}
\eea
In order to compute $N_{CR}$ one needs to know the $\beta$ evolution 
and hence the ionization
history of the IGM.  We model this using a very simple 
reionization scenario with the aim to merely compute
$N_{CR}$ in a self-consistent setting rather than addressing 
the issues related to early reionization. 
The rate of change in the number density
of ionized hydrogen (i.e. n$_{\rm H~{\sc II}}$) can be written as,
\begin{eqnarray}
\frac{d n_{HII} } { dt } &= &\dot{N}_{\gamma } + \int N_{CR}(\beta , z, z_0)
c n_{HI} \sigma_{HI} \beta d\beta\nonumber\\
 & &~ - \alpha _B n_{HII} n_e .
\label{eqreion}
\end{eqnarray}
The right hand side of Eq.~\ref{eqreion} gives the
rates of photoionization, cosmic ray ionization and recombination
respectively. We assume all the Lyman continuum photons
entering the IGM with a rate $\dot{N}_\gamma$ are used for 
ionizing H~I.
For a SFR of 1 $M_\odot$ yr$^{-1}$ about
24000 s$^{-1}$ and 4000 s$^{-1}$  Lyman continuum photons are 
generated per baryon, for ``top-heavy'' and ``Salpeter'' IMF  respectively.
This is used with an escape fraction of 0.1 to compute
$\dot{N}_\gamma$.
$\sigma _H $ is the cross-section 
for H~I ionization by protons
\cite{spitzer} and 
$\alpha_B$  is the case-B recombination coefficient.
We find CR ionization is 
sub-dominant compared to the photoionization.

We assume ``top-heavy'' IMF prior to the epoch of reionization
and ``Salpeter'' IMF after that.
The value of n$_{HII}$ at a given $z$ is obtained using
Eq.~\ref{eqreion} and the analytical fit to the observed
SFR density as a function of $z$ (Eq.~\ref{SFR}).
We get the epoch of reionization (where $n_{HII}$ = $n_H$) at 
$z_r \simeq$ 10. Note that this is independent of our choice of $\eta$
because while $R\propto \eta$ the UV escape fraction $f_e\propto 1/\eta$. 
However, as SNe rate increases with  $\eta$ the CR flux density 
will depend on $\eta$. 
Using the reionization model, and Eqs. \ref{SFR} - \ref{NCR} 
we compute the spectrum of CR at any given $z$, starting at $z_0=15$.
Such spectra 
are used below to compute the heating of the IGM due to CRs.

\begin{figure}
\centerline{\vbox{\psfig{figure=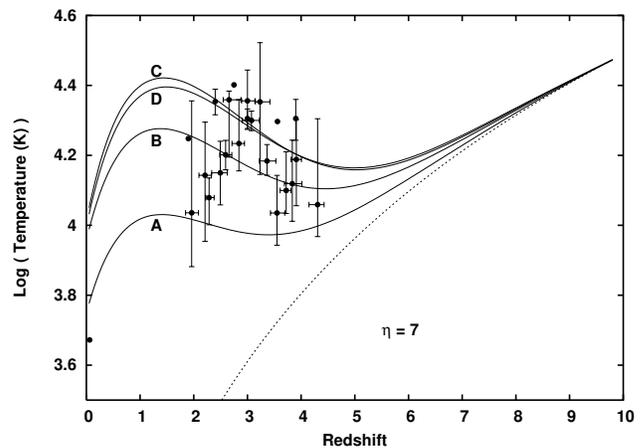,width=8.6 cm,angle=-90}}}
\caption{Predicted temperature evolution of the IGM for different
injected spectra given in Fig.~\ref{spectra}. We follow the labeling
convention as in Fig.~\ref{spectra}. Temperature evolution for the
adiabatic cooling is shown with a dotted curve.
The data points are a compendium of observations presented in \cite{schaye}.}
\label{temperature}
\end{figure}
In our model calculations we assume (i) IGM reaches an uniform 
temperature of $T = 3 \times 10^4$ K at $z = z_r$, 
(ii) heating by UV photons, Compton cooling by CMB and 
recombination cooling are negligible for $z<z_r$ and 
(iii) for simplicity the IGM is assumed to be 
of uniform density.
Thus the thermal state of the IGM  after reionization is described by,  
 \bea
\frac { d \cal E } { d z } =  \frac { 5 \cal E } { (1 + z ) } - \Gamma _{CR}
\label{epsilon}
\eea
with $\cal E $ is the thermal energy per unit volume of the IGM. The first
term in the right hand side of Eq. \ref{epsilon} gives adiabatic cooling rate.
The second term $ \Gamma _{CR} $ is the cosmic rays heating rate
given by
\ber
\Gamma _{CR}(z, z_0 )  & = & \int \frac { dE(\beta ) } { dz }
N_{CR}( \beta , z, z_0 ) d\beta   \nonumber \\
                       & = & \int \frac { dE(\beta ) } { dt } 
\frac{dt } { dz } N_{CR}( \beta , z, z_0 ) d\beta . 
\label{gammacr}
\eer
Here, $dE / dt $ is obtained from the first term of Eq. \ref{dbetadt}.
Using $ {\cal E} = 3nkT/2 $, 
where $T$ is the temperature and $n$ the number density of particles 
in the IGM, one can obtain
\ber
T  ( z ) & = & T ( z_r ) \left( \frac{ 1+ z } { 1 + z_r } \right)^{2}\times\nonumber\\ 
&&\left[ 1 + \frac {(1 + z_r ) ^2 } { 3 n_0 k T (z_r) }
 \int\limits _z ^{z_r}\frac { \Gamma _{CR}(z', z_0) } { ( 1 + z' ) ^5 } dz' \right],
\label{eqtem}
\eer
where $n_0=n_e(z=0)$.

The temperature evolution of the IGM for different injected spectra
(shown in Fig.~\ref{spectra}) computed using Eq.~\ref{eqtem} 
are shown in Fig.~\ref{temperature}.  The 
temperature of the IGM goes well below the observed range 
due to adiabatic cooling in the absence of heating sources 
(dotted curve in Fig.~\ref{temperature}). 
However for all the models discussed here,  CR heating leads to,
$T\ge10^4$K in the redshift range $2\le z \le4$. 

For case (I) the average grammage traversed by the CRs in 
the high $z$ star forming galaxies has to be less than 
$10$ g cm$^{-2}$ seen in the Galaxy, 
in order to produce the observed mean 
IGM temperature (curves A and B in Fig.~\ref{temperature}). 
A lower grammage is reasonable for the 
high $z$ galaxies, as mentioned earlier. 
These models predict cosmic ray energy density
$U_{cr} \sim 6 \times 10^{-3}$ eV cm$^{-3}$ at $z = 0$.
Models with adiabatic losses (case (II)) or termination
shocks (case (III)) produce higher temperatures,
as they efficiently inject lower energy CRs into the IGM.
At the same time they lead to overall lower
cosmic ray energy densities of the order 
$1.5\times 10^{-4}$ eV cm$^{-3}$ and 
$5.7 \times 10^{-4}$ eV cm$^{-3}$ for case (II) 
and (III) respectively at $z=0$.

Models of shock acceleration \cite{hillas} 
predict $\epsilon$ as high as $0.3-0.4$. 
In such a case even a model with grammage as seen
in our galaxy will produce higher 
temperatures than the observed mean.  
Simulations of pregalactic outflows \cite{mori}, 
find up to 30\% of the supernovae energy can go
into winds. The resulting temperature in case (III) can 
thus be much higher.
We considered CRs produced in the SNe and termination shocks, 
CRs generated in structure formation 
shocks \cite{miniati} could provide additional heating. The major 
uncertainties are related to the extent to which 
the IGM is magnetized and confinement issues.
More detailed computations in 
the framework of structure 
formation models are in progress.

We have explored the influence of 
CRs escaping from the early star forming regions on the
thermal history of the IGM. The estimated IGM temperature
evolution is sensitive to the star formation rate ($\eta$) 
and the confinement of CRs. 
Using a star formation rate density constrained by observations 
and plausible range of parameters, the observed temperature of 
the IGM at $2 \le z \le 4 $ can be explained as due to the CR 
heating, even if the epoch of reionization is as early as $z = 10$. 
This implies that low energy CRs in the IGM can erase the
thermal memory of reionization. 
In conclusion, cosmic rays do provide an important source
for heating the intergalactic medium, that can explain the 
measured IGM temperatures at high redshifts.
 
SS thanks CSIR for providing support for this work.
We thank A. Shukurov for helpful comments.

\end{document}